\documentclass[aps,prd,reprint,superscriptaddress,notitlepage,nofootinbib]{revtex4-1}

\usepackage[colorlinks=true,urlcolor=blue,linkcolor=blue,citecolor=blue]{hyperref}
\usepackage{amsmath}
\usepackage{graphicx}
\usepackage{bm}
\usepackage{subfigure}
\usepackage{amssymb}
\usepackage{cancel}
\usepackage{color}
\usepackage{listings}
\usepackage{xcolor}
\usepackage{slashed}
\usepackage{microtype}

\newcommand{\be}{\begin{equation}}
\newcommand{\ee}{\end{equation}}
\newcommand{\beq}{\begin{equation}}
\newcommand{\eeq}{\end{equation}}
\newcommand{\aln}[1]{\begin{align}#1\end{align}}
\newcommand{\bea}{\begin{eqnarray}}
\newcommand{\eea}{\end{eqnarray}}
\newcommand{\besp}{\begin{equation}\begin{split}}
\newcommand{\eesp}{\end{split}\end{equation}}

\newcommand{\Dfbd}{\mathord{\buildrel{\lower3pt\hbox{$\scriptscriptstyle\leftrightarrow$}}\over {D}_{\mu}}}

\def\0{\textbf{0}}
\def\1{\textbf{1}}
\def\2{\textbf{2}}
\def\3{\textbf{3}}
\def\4{\textbf{4}}
\def\5{\textbf{5}}
\def\6{\textbf{6}}
\def\7{\textbf{7}}
\def\8{\textbf{8}}
\def\9{\textbf{9}}

\begin{document}

\title{PBH formation from overdensities in delayed vacuum transitions}

\author{Kiyoharu Kawana}
\email{kkiyoharu@kias.re.kr}
\affiliation{School of Physics, Korea Institute for Advanced Study, Seoul 02455, Korea}

\author{TaeHun Kim}
\email{gimthcha@kias.re.kr}
\affiliation{School of Physics, Korea Institute for Advanced Study, Seoul 02455, Korea}
\affiliation{Center for Theoretical Physics, Department of Physics and Astronomy, Seoul National University, Seoul 08826, Korea}

\author{Philip Lu}
\email{philiplu11@gmail.com}
\affiliation{Center for Theoretical Physics, Department of Physics and Astronomy, Seoul National University, Seoul 08826, Korea}


\begin{abstract}
Primordial black hole (PBH) formation from first-order phase transitions (FOPTs) combines two prevalent elements of beyond the Standard Model physics with wide-ranging consequences. We elaborate on a recently proposed scenario in which inhomogeneities in vacuum energy decay seed the overdensities that collapse to PBHs. In this scenario, the PBH mass is determined by the Hubble mass as in conventional formation scenarios, while its number density is determined by the nucleation dynamics of the FOPT. We present a detailed study of the formation probability including parameter dependencies. In addition, we generate populations in the open mass window as well as for the HSC and OGLE candidate microlensing events. This mechanism inevitably creates PBHs in generic FOPTs, with significant populations produced in slow and moderately strong phase transitions.
\end{abstract}

\maketitle

\allowdisplaybreaks

\section{Introduction}

The existence of primordial black holes (PBHs) would have many interesting and diverse consequences. In the asteroid mass window $10^{-16} M_\odot \lesssim M_{\rm PBH} \lesssim 10^{-11} M_\odot$, the PBH mass fraction is free of constraints, making this mass range a popular candidate for PBH dark matter (DM)~\cite{Carr:2020xqk, Carr:2020gox, Carr:2009jm}. As a subdominant component of DM, PBHs can contribute to the LIGO-VIRGO-KAGRA population of detected binary black hole mergers~\cite{Nakamura:1997sm,Bird:2016dcv,Cotner:2016dhw,Cotner:2017tir,Raidal:2017mfl,Eroshenko:2016hmn,Sasaki:2016jop,Clesse:2016ajp,Cotner:2018vug,Cotner:2019ykd,Flores:2020drq,Flores:2021jas,Flores:2021tmc,Wang:2022nml} or seed supermassive black holes~\cite{Bean:2002kx,Kawasaki:2012kn,Clesse:2015wea}. Interestingly, two microlensing experiments HSC~\cite{Niikura:2017zjd} and OGLE~\cite{Niikura:2019kqi} have reported candidate events which may be consistent with a population of PBHs at $M_{\rm PBH}^{}\sim10^{-9} M_\odot$ and $M_{\rm PBH}^{}\sim 10^{-4} M_\odot$, respectively. PBHs are leading candidates for DM and ones with abundant astrophysical consequences.

It is therefore unsurprising that PBH-production mechanisms have saturated the literature in recent years. In addition to the classic formation from primordial overdensities~\cite{Carr:1974nx,Carr:1975qj}, models involving first-order phase transitions (FOPTs) have become popular~\cite{Crawford:1982yz,Hawking:1982ga,La:1989st,Moss:1994iq,konoplich1998formation,Konoplich:1999qq,Kodama:1982sf,Lewicki:2019gmv,Kusenko:2020pcg} (see also~\cite{Animali:2022otk} for an inflationary scenario but with metastable vacuum). Generically, FOPTs soften the equation of state of the relativistic plasma and stimulate the growth of overdensities, leading to increased PBH production~\cite{Jedamzik:1999am,Musco:2012au,Byrnes:2018clq,Jedamzik:1998hc,Jedamzik:2020omx,Carr:2019kxo,Carr:2019hud,Dolgov:2020sov,Franciolini:2022tfm,Juan:2022mir,Lu:2022yuc}. In models with additional particles, they can accumulate and form PBHs and other compact objects~\cite{Gross:2021qgx,Baker:2021nyl,Baker:2021sno,Kawana:2021tde,Marfatia:2021hcp,Huang:2022him,Davoudiasl:2021ijv,Jung:2021mku,Hashino:2021qoq,Maeso:2021xvl,Lu:2022paj,Kawana:2022fum,Lu:2022jnp,Kawana:2022lba}. 

In particular, Ref.~\cite{Liu:2021svg} calculated the abundance of FOPT-induced PBHs that result from the overdensities in the false vacuum (FV) regions with delayed decay (see~\cite{He:2022amv} for a specific FOPT model realizing the same scenario). The essence of this mechanism is the stochastic nature of the FOPT, which creates inhomogeneities in the subsequent radiation energy distribution after the vacuum transition.
These perturbations may then grow and collapse into PBHs in the standard manner. We seek to elaborate and present a more detailed investigation of this interesting scenario, providing a quantitative understanding of the production probability. 
The PBH prediction of this mechanism is sensitive to the parameters of the FOPT, so it is possible to realize any mass and energy fraction given suitable models. 
In this paper, we simply choose benchmark parameters of FOPTs that correspond to the open mass window ($\sim 10^{-15}M_\odot^{}$) and the two microlensing events mentioned above, leaving concrete model buildings for future investigation.  

In this paper, we describe the formation of PBHs from vacuum energy decay in a FOPT. In Sec.~\ref{sec:FOPT}, a general overview of FOPTs and the relevant equations are given. In Sec.~\ref{sec:formation}, we construct the formalism to calculate the PBH formation probability and abundance. In Sec.~\ref{sec:PBH}, the numerical process and results are described. Finally, in Sec.~\ref{sec:discussion}, we discuss the results and draw conclusions.

\section{First-Order Phase Transition}
\label{sec:FOPT}

We summarize the progression of a FOPT and list out key equations. In the following, we take a phenomenological approach for the FOPT parameters $\alpha$ and $\beta$ defined below. Our arguments are largely independent of the underlying model, provided that the phase transition has a strong vacuum energy $\alpha\sim \mathcal{O}(0.1)$ and slow pace $\beta/H \sim \mathcal{O}(1)$, where $H$ is the Hubble expansion rate.

In a FOPT, the Universe is initially in a stable FV state which becomes unstable below the critical temperature $T_{\rm cri}$ (see Refs.~\cite{Megevand:2016lpr,Kobakhidze:2017mru,Ellis:2018mja,Ellis:2020awk,Wang:2020jrd}) and corresponding time $t_{\rm cri}$. The nucleation rate for true vacuum (TV) bubbles is given by
\begin{equation}
\label{eq:nucl}
    \Gamma(t) = A(t) e^{-S(t)}~,
\end{equation}
where $S(t)$ is the bounce action of the three- (from thermal fluctuations) or four-dimensional (from quantum tunneling) instanton solution. $\Gamma(t)$ is the number of bubble nucleations per unit time per unit physical volume in FV regions.

The time dependence of the nucleation rate stems from the temperature dependence of the finite-temperature effective potential. As the temperature varies with time due to the expansion of the Universe and the progression of the phase transition, the effective potential and the corresponding nucleation rate $\Gamma(t)$ vary accordingly.

The qualitative behavior of the effective potential depends on whether the potential barrier separating the two vacua disappears in a finite temperature below $T_{\rm cri}$ or it persists even in the zero-temperature limit~\cite{Megevand:2016lpr}. As the Universe cools down, the nucleation rate $\Gamma$ continually increases for the former case. In the latter case, it peaks at some intermediate temperature and subsequently decreases, where only the quantum tunneling can nucleate new bubbles in the zero-temperature limit. The statistics of the postphase transition universe and the completion of the phase transition (having no FV regions) depend on the low-temperature behavior. We discuss this consistency problem in Sec.~\ref{ssec:criterion} and in more detail in Appendix~\ref{app:completion}, by stating the conditions that should be satisfied in order to have a FOPT that successfully ends while producing PBHs by the scenario considered here.

The temperature evolution is complicated by the moderate reheating from the release of latent heat and the differing Hubble rate in a two-component radiation-vacuum universe. The effects of reheating are minor in a weak FOPT with low latent heat. In a strong detonationlike FOPT, the bubble walls move at nearly the speed of light, faster than the sound speed, so the latent heat remains in TV and the nucleation rate in the FV is unaffected. An inhomogeneous Hubble rate arises from the constant energy density in the vacuum-dominated FV regions and the redshifting radiation-dominated TV regions (see Sec.~\ref{sec:formation}). However, we neglect this effect as the inhomogeneous expansion rates are not directly relevant to the focus of this paper, and instead use the average energy density to calculate the Hubble rate.

For analytical purposes, the exponential form
\begin{equation}
\label{eq:expnucl}
    \Gamma(t) \approx \Gamma_0 e^{\beta(t-t_0^{})}~,~~\beta = -\frac{dS(t)}{dt}\bigg|_{t=t_0^{}}^{}~
\end{equation}
is a good approximation when $\beta/H \gg 1$, and the second derivative of $S$ is negligible with respect to $\beta$~\cite{Megevand:2016lpr} at an arbitrary fixed time $t_0$. Hereafter, we take $t_0 = t_{\rm cri}$, the critical time in which the two vacua become degenerate. The average size of the FV bubbles is related to the bubble wall velocity $v_w^{}$ and $\beta$ by $R\sim 1.12 v_w^{}/\beta$~\cite{Kawana:2021tde,Lu:2022paj}, which implies that $\beta/H \lesssim {\cal O}(1)$ is necessary to have enough horizon-sized FV regions that can collapse into a detectable number of PBHs. For numerical calculations, we consider values of $\beta/H \sim \mathcal{O}(1)$ to marginally satisfy both these conditions.
Another key parameter is the ratio of the vacuum energy to the plasma energy density $\alpha \equiv \Delta V/\rho_{\rm SM}^{}(T_{\rm cri})$, which measures the strength of the FOPT. 
In this paper, we make an approximation by ignoring the temperature evolution of $\Delta V$ and let it be a constant after $T_{\rm cri}$. This corresponds to a situation where only the height of the potential barrier varies, giving Eq.~\eqref{eq:expnucl}, while the vacua energies remain the same. Also, we simply focus on $\alpha\lesssim \mathcal{O}(1)$ to avoid a long-duration second inflationary phase.

As the phase transition progresses, the volume fraction of the FV region shrinks from $f_{\rm fv} = 1$ at $T>T_{\rm cri}^{}$ to $f_{\rm fv} \xrightarrow{} 0$ at $T\ll T_{\rm cri}^{}$, and can be computed by
\begin{equation}
\label{eq:ffv}
    f_{\rm fv}(t) = \exp \left[ -\frac{4\pi}{3} \int^{t}_{t_{\rm cri}} dt' \, a^3(t') \, \Gamma (t') \, r^3(t, t') \right]~,
\end{equation}
where 
\begin{equation}
\label{eq:radius}
	r(t, t') = v_w \int^t_{t'} dt'' \frac{1}{a(t'')} ~
\end{equation}
is the comoving size of a bubble nucleated at $t'$. 
Under the constant scale factor approximation or constant Hubble rate approximation during the transition and using the exponential approximation (\ref{eq:expnucl}), $f_{\rm fv}^{}(t)$ can be analytically solved to be~\cite{Turner:1992tz}
\begin{eqnarray}
\label{eq:ffvexp}
        f_{\rm fv}^{}(t) &=& \exp\left[-I_{\ast} e^{\beta(t-t_{\rm cri}^{})}\right], \nonumber \\
        I_{\ast}&=&
        \begin{cases}
		  \frac{8\pi v_w^3 \Gamma_{\rm cri}}{\beta^4} & \text{(const $a$),}\\
            \frac{8\pi v_w^3 \Gamma_{\rm cri}}{\beta (H+\beta) (2H+\beta) (3H+\beta)} & \text{(const $H$).}
		\end{cases}
\end{eqnarray}
Note that we have approximated $t_{\rm cri}^{}\rightarrow -\infty$ when $H=$ const. Here we see that there are competing effects for PBH production. As detailed in Sec.~\ref{sec:formation}, the constant vacuum energy $\Delta V$ grows in strength to the surrounding plasma at lower temperatures and later times. Conversely, the FV fraction steadily decreases with increasing time, constraining the PBH production to peak at intermediate times.

\section{PBH Formation Mechanism}
\label{sec:formation}
In the classical analytic prescription for PBH formation, a PBH forms when horizon-sized perturbations exceed a critical overdensity threshold $\delta>\delta_c \sim 0.45$~\cite{Musco:2020jjb}. The spectrum of curvature perturbations is usually generated during inflation and reenters the horizon after reheating. Here, we consider the scenario outlined in Ref.~\cite{Liu:2021svg} in which overdensities are generated stochastically during a FOPT. 
However, we investigate the formation mechanism in more detail and comment on the different methodologies in this paper and Ref.~\cite{Liu:2021svg} in Sec.~\ref{ssec:compare}.

For a moderately strong FOPT $\alpha \sim \mathcal{O}(0.1 \text{--} 1)$, the vacuum energy is a significant component of the overall energy density which even temporarily keeps the Hubble rate $H(t)$ constant.  
This vacuum energy will seed the fluctuations when converted to radiation by the phase transition. Because of the conservation of entropy, the background plasma radiation density redshifts with scale factor $a$ as $\propto a^{-4}$, but the vacuum energy density $\Delta V$ stays constant. 
Therefore, regions of FV will have proportionally larger energy densities with increasing time. If the resultant density contrasts are large enough, they gravitationally collapse and PBHs are formed (see Sec.~\ref{ssec:criterion} for detailed PBH formation criterion), whereas deficient overdensities will decay and disappear soon due to the radiation pressure. 

In the paper, we focus on the PBHs from horizon-sized overdensities only and discard the subhorizon ones, since the formation criterion for these are not well understood.

\subsection{Formation criterion} \label{ssec:criterion}
Here the quantitative criterion for PBH formation is stated. First, we define a time $t_{1.45}^{}$ at which the local energy density of a FV point is $1+\delta_c^{} \simeq 1.45$ of the background average density,
\begin{equation}
\label{eq:t145}
   1+\delta:=\frac{\Delta V + \rho_{\rm SM}^{}(t_{\rm 1.45})}{\Bar{\rho}(t_{\rm 1.45})} = 1.45~,
\end{equation}
where $\rho_{\rm SM}^{}(t)$ is the Standard Model (SM) plasma energy density that was present before the transition, and $\Bar{\rho}(t)$ is the average total energy density of the Universe (see Sec.~\ref{ssec:evolution} for more concrete definitions).

Then we use a simple but justified criterion for PBH formation: Any Hubble volume that is entirely covered by FV after time $t_{\rm 1.45}$ eventually collapses to a PBH. There are a few subtleties excluded by this prescription. First, the value of $\delta_c = 0.45$ is applicable only to radiation-dominated plasma and should be scaled according to its effective equation-of-state parameter $w$~\cite{Musco:2012au} which would be different due to the presence of vacuum energy. Furthermore, the overdensity $\delta$ cannot accumulate and collapse when the horizon is dominated by the FV until it decays into radiation, delaying the growth phase of the perturbation past $t_{1.45}$. However, this delayed growth is offset by the relative increase in the FV energy $\Delta V$ to the surrounding plasma during this time. These omitted effects tend to cancel and we leave a more in-depth investigation of PBH formation involving numerical simulations to future work. 

The horizon-sized FV regions will expand exponentially if they remain intact. This resembles the old inflation scenario, which was abandoned due to the prediction that inflation never ends because the vacuum transition cannot overcome the exponential expansion of FV regions. However, there is one important difference for FOPT: The nucleation rate is not restricted to be much smaller than the Hubble scale and also allowed to increase exponentially as in Eq.~(\ref{eq:expnucl}). So for our scenario to be consistent with the successful completion of a FOPT, the nucleation rate should catch up with the volume expansion rate, making the physical volume of FV $\propto f_{\rm fv}(t) \, a^3(t)$ decrease~\cite{PhysRevD.46.2384,Kawana:2022fum}. More detailed conditions to achieve this are explained in Appendix~\ref{app:completion} with a numerical demonstration. We are assuming one of the three cases with exponential approximation (\ref{eq:expnucl}) for the nucleation rate; see the Appendix for more details. Any deviation from our assumption will result in different quantitative results, but the physics of PBH formation is not altered.

\subsection{Formation probability}
\label{ssec:formprob}

\begin{figure}[t] \centering 
\includegraphics[width=0.48\textwidth]{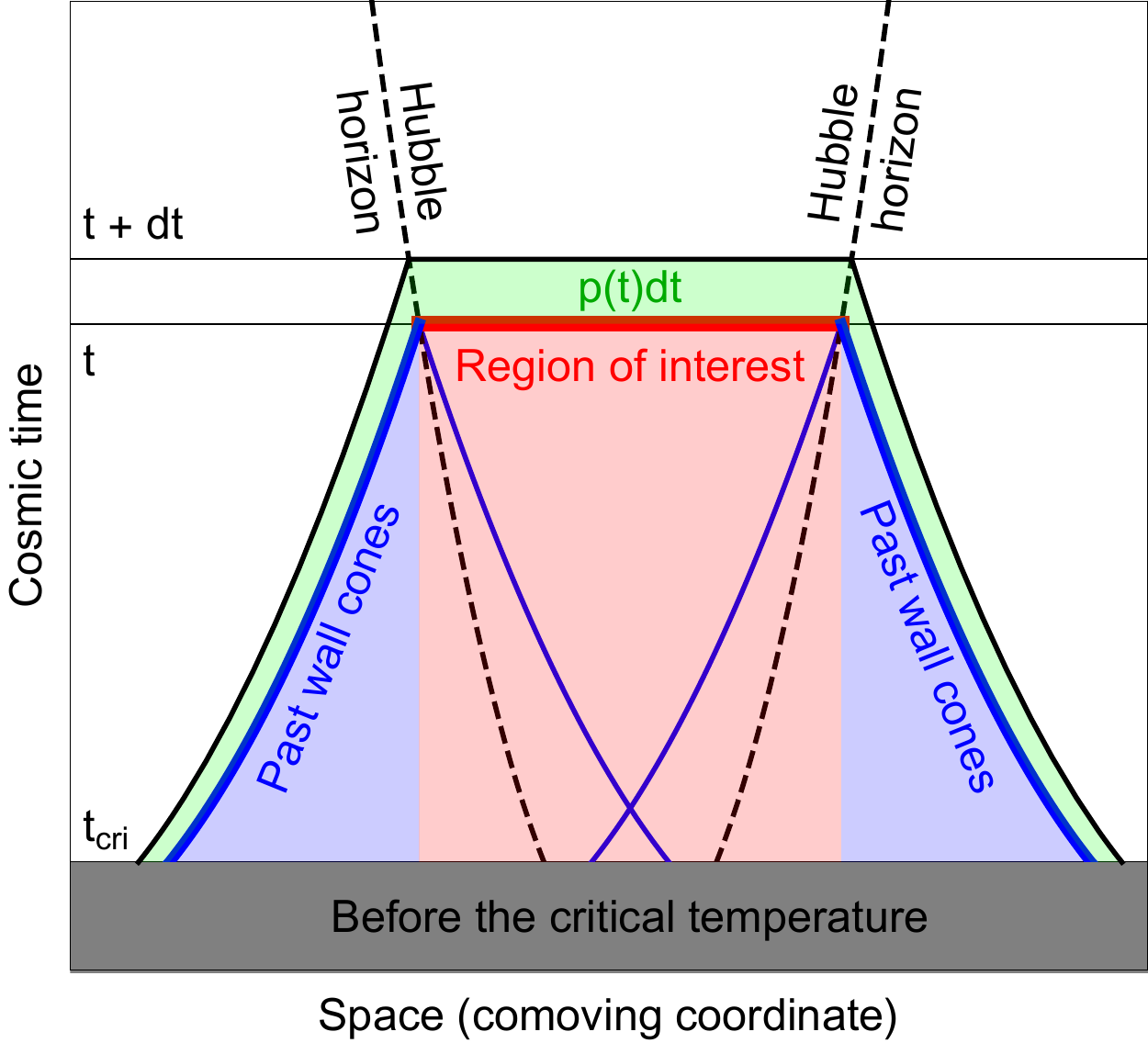}
\caption{
Wall cone diagram for PBH formation. A nucleation in either the horizon volume (red) or the causative past wall cone (blue) would trigger the phase transition. A nucleation in the green spacetime slice would induce the phase transition between time $t$ and $t+dt$.
} 
\label{fig:wallconeQ2Q3}
\end{figure}

We calculate the time-dependent probability $p(t)$ of PBH formation per unit time for any given Hubble volume. We then apply this in Sec.~\ref{sec:PBH} to find the PBH mass spectrum and abundance.

In the following, we use the wall cone formalism developed in Ref.~\cite{Lu:2022paj} with the comoving radius of a TV bubble given by Eq.~\eqref{eq:radius}. The wall cone formalism describes the propagation of bubble walls as cones in spacetime, corresponding to light cones only with light speed $c$ replaced by wall velocity $v_w$. A future wall cone coming out from a nucleation point represents the region of TV, while the past wall cone of a spacetime point represents the region for nucleation to put that point in TV.

For a Hubble volume to be entirely in FV at time $t$, (i) there can be no nucleation within the Hubble volume for time $t'<t$ and (ii) no nucleation within the past wall cone of the Hubble volume. 
A nucleation outside the horizon at time $t'<t$ can propagate to the Hubble volume if it is less than a distance $r(t,t')$ away.
In Fig.~\ref{fig:wallconeQ2Q3}, we display these three regions in spacetime. 
The probability $p(t) dt$ represents the transition probability of the Hubble volume between $t$ and $t+dt$. 
We now derive this probability. To streamline the discussion, we use comoving quantities, with the comoving nucleation rate defined as
\begin{equation}
\label{eq:gammcom}
	\Gamma_{\rm com}(t) = \Gamma(t) a^3(t).
\end{equation}
In an infinitesimal time interval $dt$, the probability of no nucleation in an infinitesimal comoving volume $dV_{\rm com}$ is $1-V_{\rm com} \Gamma_{\rm com} dt$. Summing infinitesimal volumes and taking the limit, the probability of no nucleation within a given comoving volume until the time $t_*^{}(\vec{x})$ for each comoving point $\vec{x}\in V_{\rm com}^{}$ is
\begin{eqnarray}
	P_{\rm \slashed{n}}(V_{\rm com}; t_*) &=& \prod_i [1-V_{\rm com} \Gamma_{\rm com} (t_i) \, dt] \nonumber \\	
	&=&\exp \left[-\int_{V_{\rm com}} d^3 \vec{x} \int^{t_*(\vec{x})}_{t_{\rm cri}}  a^3(t) \, \Gamma(t) \, dt\right]. \quad \ \label{eq:P!n}
\end{eqnarray}
We apply this formula to the transition probability $p(t)$ of a given Hubble volume by using the comoving radius from its center
\begin{equation}
	r_{x}^{} := |\vec{x}| \qquad \text{and} \qquad r_H^{}(t) := \frac{1}{a(t) H(t)}~.
\end{equation}
First, each point within the Hubble volume $r_x^{} < r_H^{}(t)$ should remain in the FV, implying $t_*(\vec{x})=t$. 
This is shown by the red region in Fig.~\ref{fig:wallconeQ2Q3}. 
On the other hand, outside the horizon, the past wall cone has ($t_*^{}<t$)
\begin{eqnarray}
	r_x(t_*^{}) = r_H^{}(t) + v_w \int^{t}_{t_*} \frac{1}{a(t)} dt :=r_H^{}(t)+r(t,t_*^{})
	\label{eq:rxt*}
\end{eqnarray}
where we assume a constant wall velocity $v_w\leq1$, and the maximum comoving radius is denoted by $r_{\rm max}^{}:=r_x^{}(t_{\rm cri}^{})= r_H(t)+r_{}(t,t_{\rm cri}^{})$. 
The corresponding outside region is represented in blue in Fig.~\ref{fig:wallconeQ2Q3}.
If we denote the inverse of Eq.~(\ref{eq:rxt*}) as $t_*(r_x^{})$, each outside point $\vec{x}$ must remain in the FV until $t_*^{}(r_x^{})$. 
Then, the survival probability of the Hubble volume, or the probability of having no TV up to time $t>t_{\rm cri}^{}$, is given by the product of the probabilities of having no nucleation in the horizon itself or in its past wall cone as
\aln{
    P (t) = &\exp \left[-\frac{4\pi}{3} \int^{t}_{t_{\rm cri}} \frac{a^3(t')}{a^3(t)} H^{-3}(t) \, \Gamma(t') \, dt'\right] \nonumber \\
	&\times \exp \left[ -4\pi \int^{r_{\rm max}}_{r_H(t)} d r_x^{} \, r_x^2  \int^{t_*^{}(r_x)}_{t_{\rm cri}} a^3(t') \, \Gamma(t') \, dt' \right]. 
	\label{eq:survival}
}
The complement $1-P(t)$ is simply the cumulative distribution function of $p(t)$, so the desired transition probability is given by
\begin{equation}
	p(t) = -\frac{d}{dt} P(t)~. \label{eq:p}
\end{equation}
As previously mentioned, we limit the domain of $p(t)$ to $t>t_{1.45}$ so that $p(t)$ can be interpreted as the probability density for the onset of PBH collapse. The mass distribution of PBH is determined by the temporal distribution of formation, as the horizon size and the overdensity $\delta$ vary with time. The mass of the PBH is related to the horizon mass at the transition time by
\begin{equation}
\label{eq:mass}
    M(t) = \gamma M_H(t) = \gamma \frac{4\pi}{3} \bar{\rho}(t) H^{-3}(t)~,
\end{equation}
where we choose the numerical prefactor $\gamma=1$, and $\bar{\rho}(t)$ is the average energy density (combined radiation and vacuum). If another value of $\gamma$ or a different PBH mass relation (e.g., critical collapse; see Sec.~\ref{ssec:numerical}) is used, the mass distribution is simply shifted with little consequence for our purposes.
$p(t)$ can be converted to the standard PBH formation parameter $\beta(M)$, which is the fraction of PBH abundance in the Universe at the formation time~\cite{Carr:2020xqk, Carr:2020gox, Carr:2009jm}. 
The (physical) number density at the formation of new PBHs for a time interval between $t$ and $t+dt$ is
\begin{equation}
    dn_{\rm PBH} = p(t) \left(\frac{4\pi}{3} \frac{1}{H^3(t)} \right)^{-1} dt
\end{equation}
giving 
\begin{eqnarray}
    \beta(M) &=& \frac{1}{\rho_{\rm SM}^{}(T)}\frac{d\rho_{\rm PBH}^{}}{d\log M} =\frac{4M}{3 T}\frac{1}{s(T)}\frac{dn_{\rm PBH}^{}}{d\log M} \nonumber \\
    &=& \frac{M H^3(t)}{\pi T} \frac{p(t)}{s(T)} \left(\frac{d \log M}{dt}\right)^{-1}, \label{eq:betaM}
\end{eqnarray}
where $s(T)$ is the entropy density of SM plasma, and both $t$ and $T$ on the rhs are understood as functions of $M$ via Eq.~(\ref{eq:mass}).
Finally, the PBH energy fraction $f_{\rm PBH}(M)$ is given in terms of $\beta(M)$ \cite{DeLuca:2020agl} as
\begin{equation}
\label{eq:fPBH}
    f_{\rm PBH}(M) = \frac{1}{\Omega_{\rm DM}} \left(\frac{M_{\rm eq}}{M}\right)^{1/2}\beta(M)~,
\end{equation}
with $\Omega_{\rm DM} =0.264$ and the horizon mass at matter-radiation equality $M_{\rm eq}=2.8\times10^{17} M_\odot$.

\subsection{Comparison with 2106.05637}
\label{ssec:compare}

\begin{figure}[t] \centering 
\includegraphics[width=0.48\textwidth]{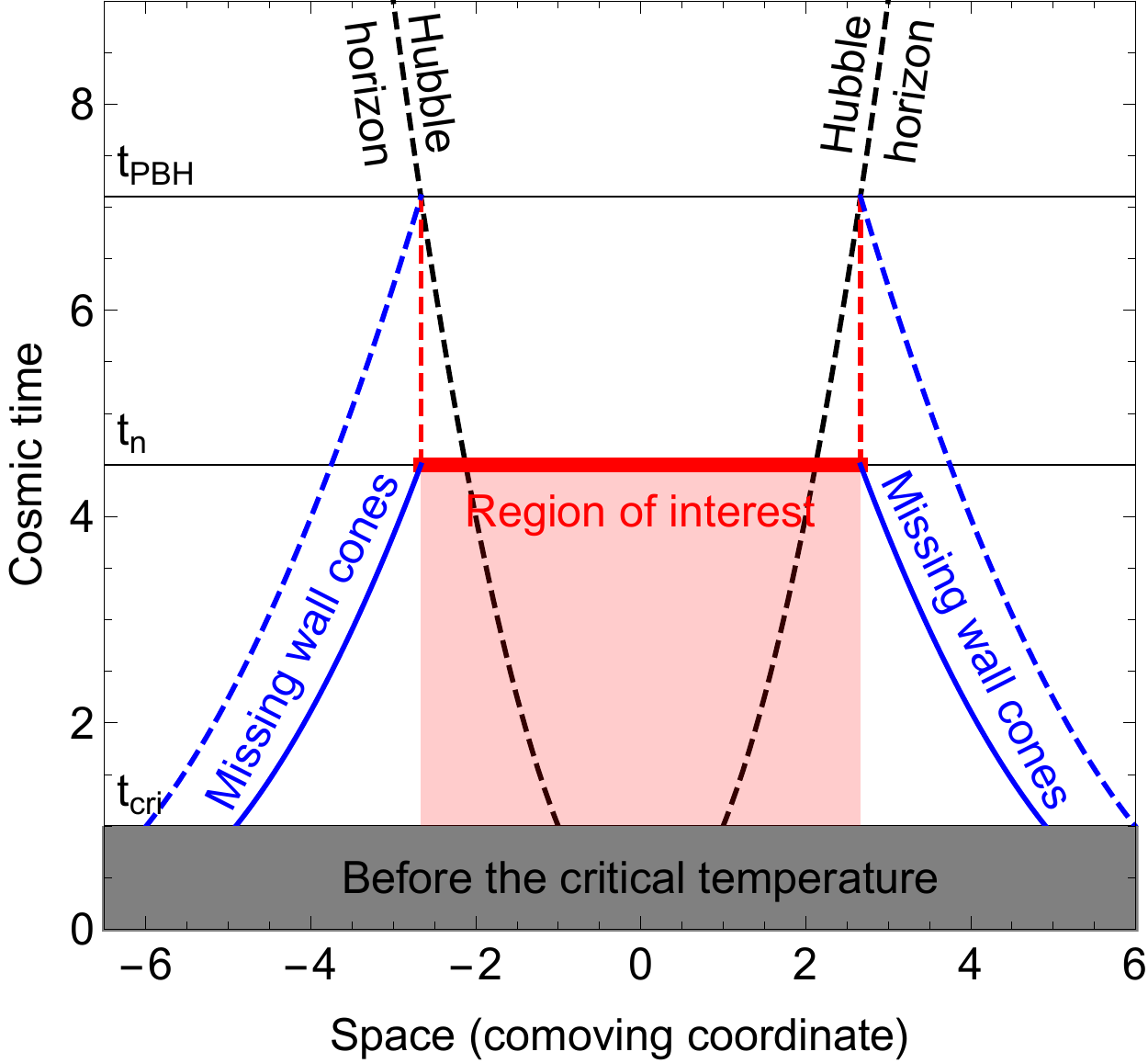}
\caption{
The underlying spacetime diagram for Refs.~\cite{Liu:2021svg, He:2022amv} with the missing wall cones, which should have been somewhere between the solid and the dashed blue curves.
} 
\label{fig:wallcone2021JLiu}
\end{figure}

In Liu \textit{et al}. 2021~\cite{Liu:2021svg}, the authors proposed a PBH formation scenario from vacuum energy during a FOPT and provided some estimates of the formation probability and PBH statistics. We elaborate on this mechanism and present improved calculations in this paper using the wall cone approach of Ref.~\cite{Lu:2022paj}. We find that Refs.~\cite{Liu:2021svg, He:2022amv} overestimate the PBH abundance. In contrast, although our methods also use analytic simplifications and contain uncertainties, they are more sophisticated and tend to conservatively estimate the formation probability, leading to a lower bound on the PBH abundance. Here we also try to clarify certain points that were implied in the qualitative exposition of Refs.~\cite{Liu:2021svg, He:2022amv}.

Although a direct comparison is difficult, the primary difference between these two methods is the past wall cone region that we consider. Our interpretation of the spacetime diagram for Refs.~\cite{Liu:2021svg, He:2022amv} is shown in Fig.~\ref{fig:wallcone2021JLiu}. Their condition for PBH production requires only delayed nucleation in the Hubble horizon itself, treating each Hubble volume as a separate universe and neglecting TV propagation from the surrounding wall cone. We find numerically that the wall cone region contributes an exponential suppression factor comparable to the contribution from the Hubble volume itself. This would result in much lower production probability and smaller $\beta(M)$, although production is also exponentially sensitive to FOPT parameters $\alpha$ and $\beta$ which can be changed to compensate for the extra suppression. Thus, we conclude that the production mechanism proposed in Ref.~\cite{Liu:2021svg} is valid, but we present a modified version and provide some details that we found missing.

\section{PBH Abundance Calculation
}
\label{sec:PBH}

We numerically evolve the FOPT equations to find the formation probability $p(t)$ and energy fraction $\beta(M)$ or $f_{\rm PBH}^{}(M)$. 
We describe the generic numerical procedure in Sec.~\ref{ssec:evolution} and show the results in Sec.~\ref{ssec:numerical}.

\subsection{FOPT evolution}
\label{ssec:evolution}

During a FOPT, the Universe has two energy components: the vacuum energy of FV $\Delta V$ and the radiation energy $\bar{\rho}_R^{}(t)$. The FOPT starts at the critical temperature $T_{\rm cri}$. At this moment, there are only homogeneous SM plasma with energy density $\frac{\pi^2}{30} g(T_{\rm cri}) T_{\rm cri}^4$ and the vacuum energy with density $\alpha$ times the former.
As the transition progresses, the vacuum energy is converted into kinetic energy for the bubble walls and delivers latent heat into the plasma. 
The average total energy density is then expressed by
\begin{equation}
\label{eq:rhototavg}
    \bar{\rho}(t) = \bar{\rho}_V^{} (t) + \bar{\rho}_R^{}(t)~,
\end{equation}
where 
\begin{equation}
\label{eq:rhovavg}
    \bar{\rho}_V^{} (t) = \Delta V \times f_{\rm fv}^{}(t)
\end{equation}
is the average vacuum energy density for the FV volume fraction given by Eqs.~\eqref{eq:ffv} and \eqref{eq:radius}, and
\begin{equation}
    \bar{\rho}_R^{}(t) = \rho_{\rm SM}(t) + \bar{\rho}_{\rm heat} (t) + \bar{\rho}_{\rm wall} (t) \label{eq:rhoR}
\end{equation}
is the total average radiation energy density, the sum of the SM plasma from the reheating after inflation, the latent heat, and the bubble wall energy. Hereafter, the wall energy is included in $\bar{\rho}_R^{}(t)$ as it is subject to the same redshift dependence of $\propto a^{-4}$~\cite{Liu:2021svg} and will release gravitational waves after collisions. $\bar{\rho}_R^{}(t)$ evolves as
\begin{equation}
\label{eq:rhoravg}
	\frac{d \bar{\rho}_R}{dt} + 4H\bar{\rho}_R^{} = -\frac{d \bar{\rho}_V^{}}{dt}~,
\end{equation}
where the converted vacuum energy goes into the last two terms in Eq.~(\ref{eq:rhoR}), and the Hubble rate $H(t)$ is determined by the Friedmann equation
\begin{equation}
\label{eq:Hubble}
	H^2(t) = \frac{\bar{\rho} (t)}{3M_P^2}~.
\end{equation}
To simplify our calculations, we assumed $H(t)$ to be homogeneous, regardless of the local composition of energy contents hence neglecting the backreaction of the inhomogeneous vacuum transition to the expansion rate. Also, we use the exponential form of $\Gamma(t)$, Eq.~\eqref{eq:expnucl}, which is marginally valid for our parameter choice of $\beta / H \gtrsim 1$. 
For even slower phase transitions, which are conducive to forming horizon-sized bubbles, the Gaussian approximation should be used~\cite{Megevand:2016lpr}; however, this may suffer from the consistency problem as explained in Appendix~\ref{app:completion}.

For a given set of FOPT and cosmological parameters, we first perform a numerical evolution of the FOPT to calculate $t_{1.45}$, i.e., Eq.~\eqref{eq:t145}. Then, using Eqs.~(\ref{eq:survival}) and (\ref{eq:p}), the transition probability of PBH formation $p(t)$ for $t > t_{1.45}^{}$ is calculated. The mass and energy fractions of the corresponding PBH are estimated by Eqs.~(\ref{eq:mass}), (\ref{eq:betaM}), and (\ref{eq:fPBH}).

\subsection{Numerical results} \label{ssec:numerical}

\begin{table}[t]  
\centering 
\begin{tabular}{|c|c|c|c|c|c|}  
\hline 
Model & $T_{\rm cri}$ & $\Gamma_{\rm cri}^{1/4}$ & $\beta/H$ & $\alpha$ & $v_w$  \\
\hline
A & $3.7\times10^{6}$ GeV & $5.39\times 10^{-6} \, \text{GeV}$ & 2.5 & 1.0 & 1   \\
\hline
B & $3.7\times10^3$ GeV & $5.47\times 10^{-12} \, \text{GeV}$ & 2.5 & 1.0 & 1   \\
\hline
C & $39$ GeV & $5.91\times 10^{-16} \, \text{GeV}$ & 2.5 & 1.4 & 1   \\
\hline

\end{tabular}
\caption{FOPT parameters for models A (PBH mass window), B (HSC), and C (OGLE) in Fig.~\ref{fig:f(M)}. $\Gamma_{\rm cri}$ and $\beta$ are for Eq.~(\ref{eq:expnucl}) with $t_0 = t_{\rm cri}$, and $\beta/H$ and $\alpha$ are evaluated at $t_{\rm cri}$. Model A is also used in Figs.~\ref{fig:rhos} and~\ref{fig:p} to demonstrate generic features of the phase transition and PBH formation.}
\label{tab:params}
\end{table}

As benchmarks, we chose three phenomenological FOPT models summarized in Table~\ref{tab:params}.\footnote{While the nucleation rate starts with $0$ at $T_{\rm cri}$ \cite{Megevand:2016lpr}, we extend the exponential approximation (\ref{eq:expnucl}) to this point for numerical calculation and take $\Gamma_{\rm cri} / H_{\rm cri}^4 \ll 1$. This value has only little effect on the results.} Model A assumes a high-temperature phase transition resulting in light mass PBHs in the mass window for DM. The other two happening at lower temperatures can produce heavier mass PBHs, which can be responsible for the candidate events of the HSC and OGLE experiments.

\begin{figure}[t] \centering 
\includegraphics[width=0.48\textwidth]{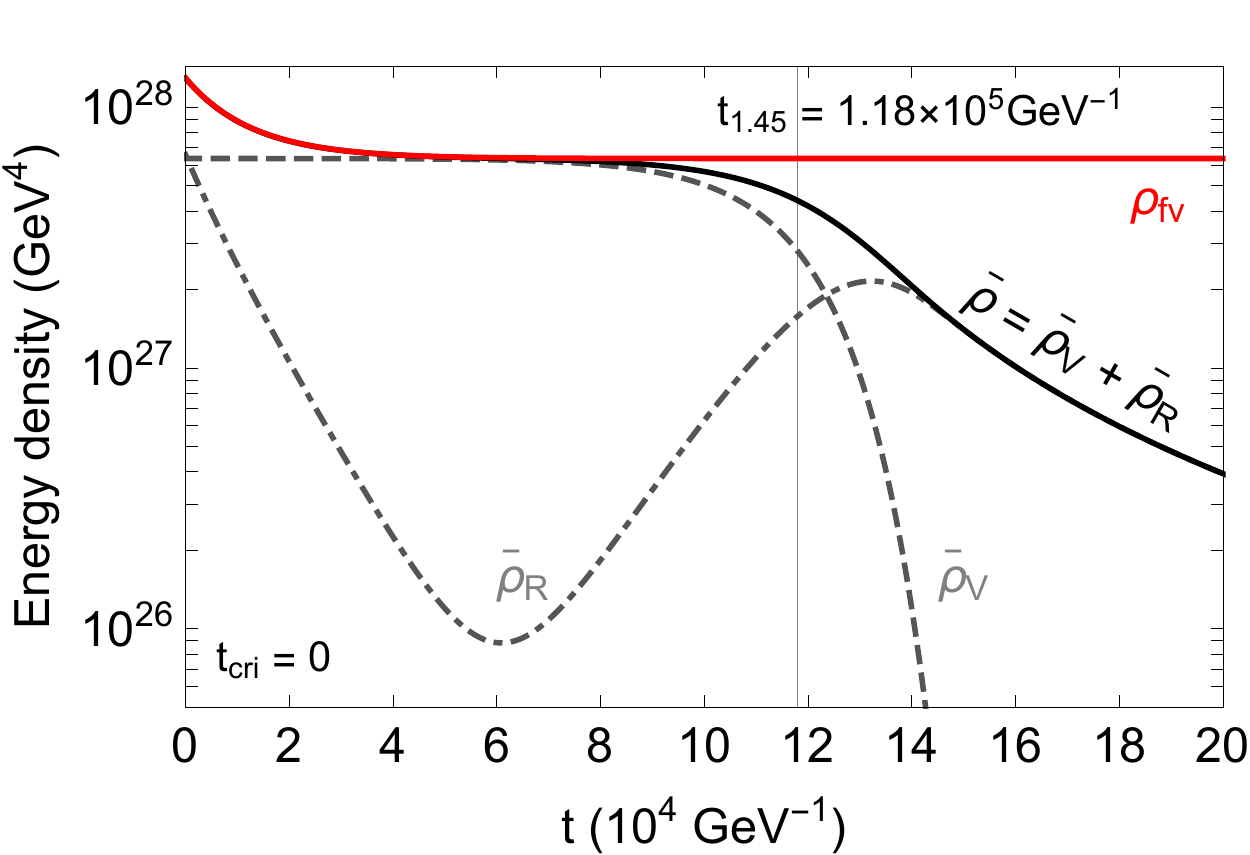}
\caption{
The average radiation density $\bar{\rho}_R(t)$, average FV energy density $\bar{\rho}_V$, and average total density  $\bar{\rho}(t)$ as functions of time. $\rho_{\rm fv}^{}(t)$ is the local density of unperturbed FV regions. 
} 
\label{fig:rhos}
\end{figure}

We first display the energy density evolution of model A in Fig.~\ref{fig:rhos}, whose behavior is generic to all three models. The average radiation energy density $\bar{\rho}_R^{}(t)$ initially falls with increasing scale factor but is subsequently heated up by the decaying FV.\footnote{Figure~\ref{fig:rhos} is for the average densities. Locally, FV regions are not reheated as discussed above Eq.~\eqref{eq:expnucl}. The nucleation rate there thus can still be approximated by the exponential form, except for the common deviation by the actual potential shape outlined in Appendix~\ref{app:completion}.} The energy density in the FV $\rho_{\rm fv}^{}$ becomes dominated by $\Delta V$ as the radiation component redshifts away. During the phase transition, the FV regions become significantly overdense $\delta>\delta_c$ with respect to the average density at $t_{1.45}^{}$. 

\begin{figure}[t] \centering 
\includegraphics[width=0.48\textwidth]{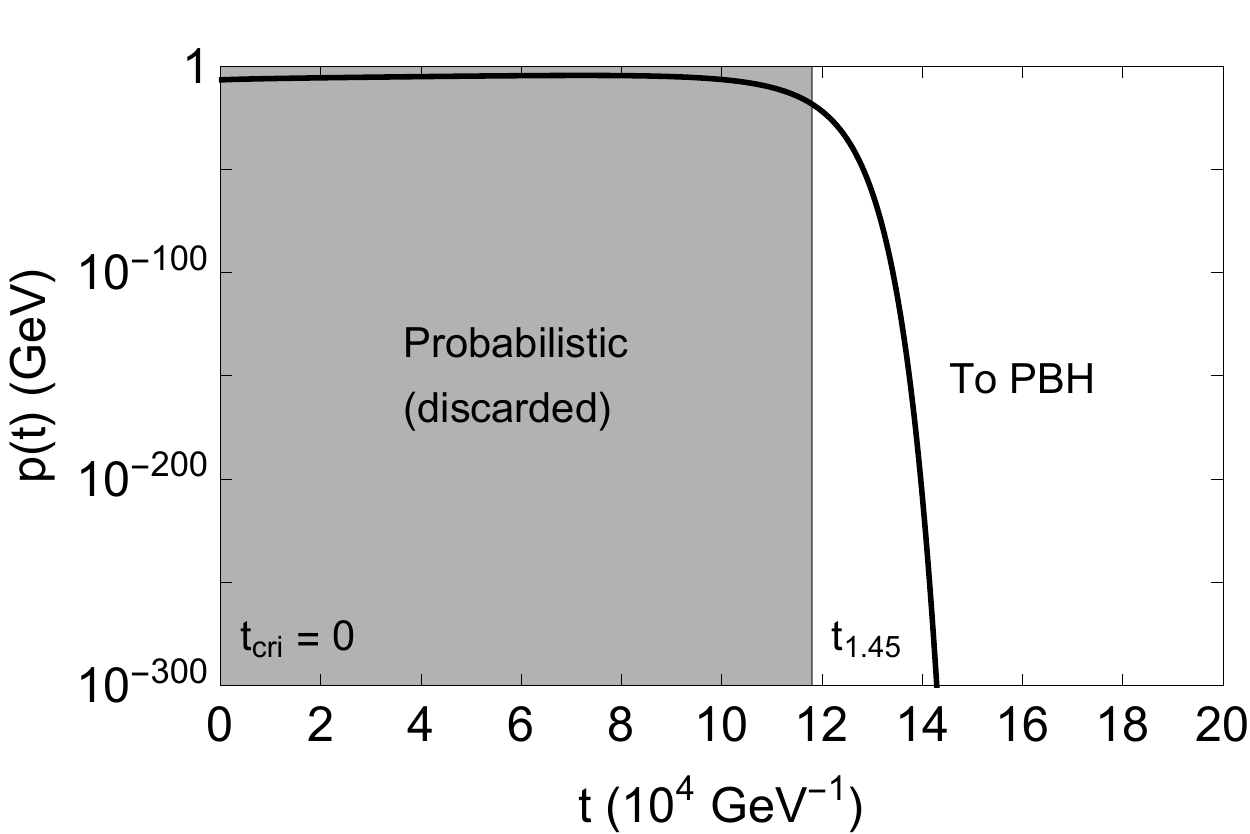}
\includegraphics[width=0.48\textwidth]{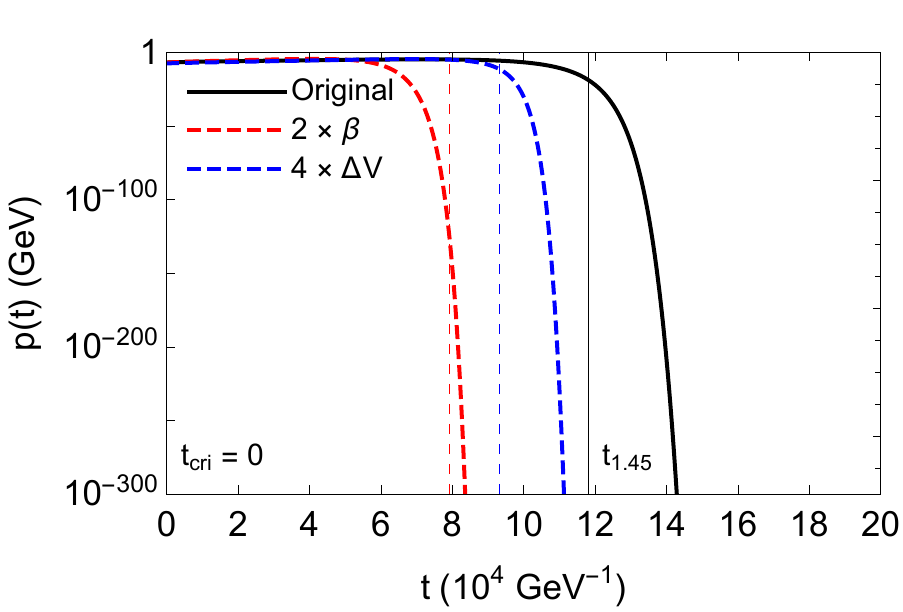}
\caption{
Top: the transition probability $p(t)$ for a Hubble volume as a function of time in model A. 
Transitions in the shaded region can only form PBHs under favorable circumstances, whereas transitions that occur after $t_{1.45}$ in the unshaded region are significantly overdense and highly likely to form PBHs. Bottom: variations of $p(t)$ and $t_{1.45}^{}$ for faster (red) or stronger (blue) FOPT parameters.
} 
\label{fig:p}
\end{figure}

If the onset of the phase transition in a Hubble volume occurs after this time, then the overdensity there will tend to grow until the FV regions are depleted so that PBH formation is almost certain. We show the PBH-producing tail of the probability density distribution $p(t)$ in Fig.~\ref{fig:p}, and the discarded region where PBH formation is unlikely. It is possible for some Hubble volumes near the boundary that begin transitioning at $t\lesssim t_{1.45}^{}$ to have $\delta>\delta_c^{}$ by the time the phase transition culminates. However, we discard these ``probabilistic" regions to provide a conservative estimate on the PBH abundance. 

In the bottom panel of Fig.~\ref{fig:p}, we show the effect of changing the FOPT parameters $\beta$, which determines the timescale of the FOPT, and $\alpha$ or the vacuum energy density $\Delta V$ (with $\beta/H_{\rm cri}$ fixed). Faster phase transitions (larger $\beta$) result in earlier TV nucleation and exponentially suppressed PBH production. Alternatively, the average size of the FV pockets is inversely proportional to $\beta$~\cite{Kawana:2021tde,Lu:2022paj}, so that there are fewer horizon-sized ones. Naturally, stronger phase transitions (larger $\Delta V$) result in larger overdensities between FV regions with energy density $\rho_{\rm fv}$ and the background $\bar{\rho}$ facilitating PBH formation. All these effects exponentially impact $p(t)$ and dominate the final PBH abundance.

\begin{figure}[t] \centering 
\includegraphics[width=0.48\textwidth]{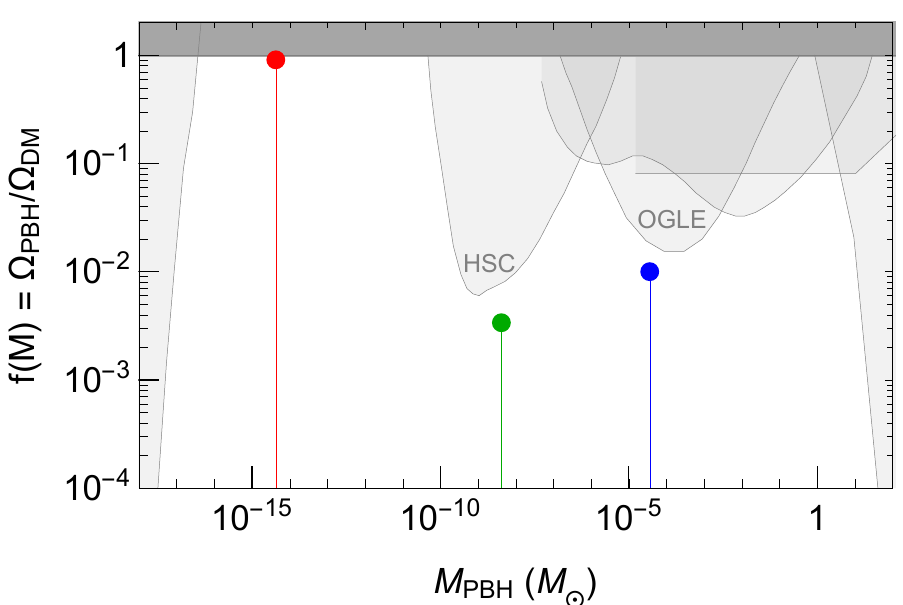}
\caption{
The PBH mass function $f_{\rm PBH}(M)$. The excluded regions by existing bounds are shown by the shaded regions. $H_0 = 67.66 \text{ km/s/Mpc}$ and $\Omega_{\rm DM} = 0.264$ are used. The three colored dots represent the abundance of PBHs in the Window (red), HSC (green), and OGLE (blue) models. The extended mass functions are sharply peaked (see Fig.~\ref{fig:p}), so we represent them as points.
} 
\label{fig:f(M)}
\end{figure}

We show the PBH distribution of the three models in Fig.~\ref{fig:f(M)}. The survival probability drops sharply after $t_{1.45}^{}$, and the scaling $\ln p(t) \sim -e^{\beta (t-t_{\rm cri})}$ can be obtained from Eq.~\eqref{eq:lnPHV} with $t_{1.45}$ replaced by $t$ and Eq.~\eqref{eq:p}. We have numerically confirmed this behavior in Fig.~\ref{fig:p}. Combined with the mass relation Eq.~(\ref{eq:mass}), the resulting PBH distributions are sharply peaked and resemble Dirac delta functions ~$\sim \delta (M_{\rm PBH}-M(t_{1.45})^{})$. However, the actual mass function would be slightly broadened to lower mass ranges if critical collapse is considered \cite{Choptuik:1992jv,Koike:1995jm,Niemeyer:1997mt}, modifying Eq.~\eqref{eq:mass}. The variation of $\delta$ of a Hubble volume after the transition time depending on the internal progression of phase transition would be responsible for it.

\begin{figure}[t] \centering 
\includegraphics[width=0.48\textwidth]{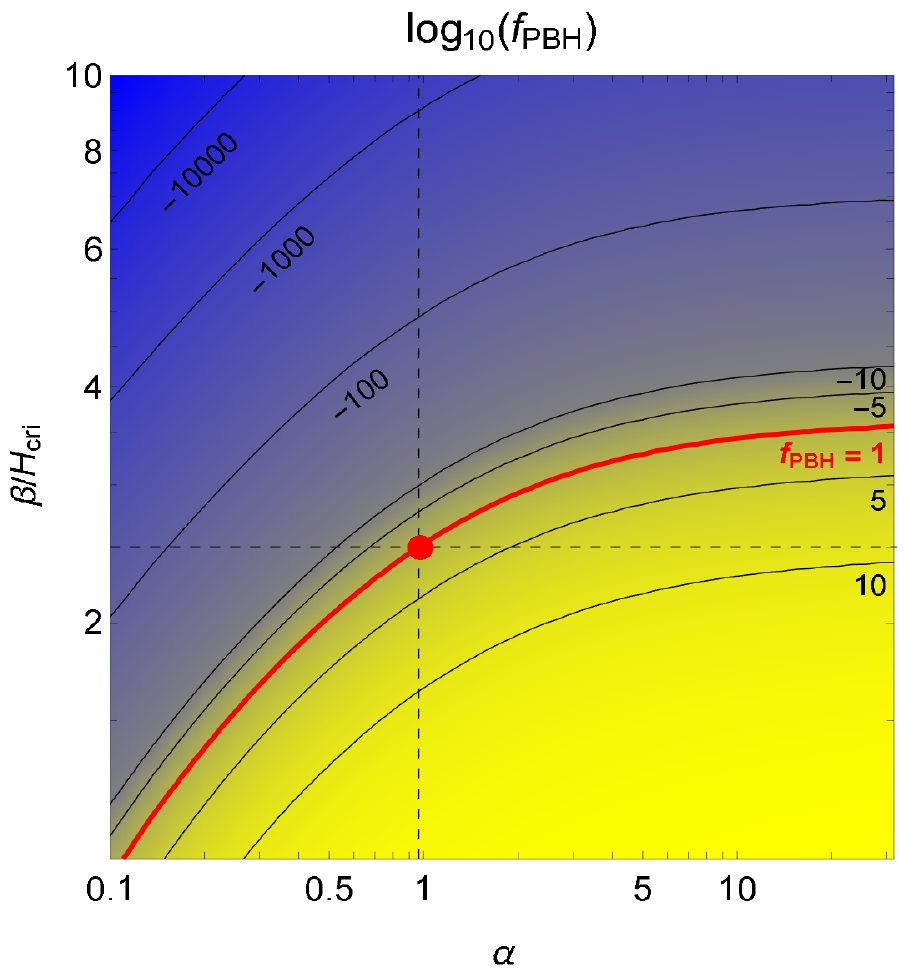}
\caption{
Contours of the final PBH abundance as a function of the FOPT parameters $\alpha$ and $\beta$, with $T_{\rm cri}$ and $\Gamma_{\rm cri} / H_{\rm cri}^4$ of model A (red dot). $\beta/H_{\rm cri}$ is fixed when $\alpha$ is varied. The red curve is where the PBHs from the FOPT constitute the full DM.}
\label{fig:paramdep}
\end{figure}

We further explore the parameter space of $\alpha$ and $\beta$ and show the PBH abundance in Fig.~\ref{fig:paramdep} using model A as a reference. As a whole, we see that the abundance increases with $\alpha$ and decreases with $\beta$, as expected from the behavior of $p(t)$ in Fig.~\ref{fig:p} and also the physical considerations. The dependence is stronger for $\beta$ than $\alpha$, and the $\alpha$ dependence is even more suppressed for $\alpha \gtrsim 10$. The $\alpha$ dependence becomes saturated by resulting in the same vacuum-dominated universe, hence the same FOPT progression for fixed $\Gamma_{\rm cri}/H_{\rm cri}^4$ and $\beta/H_{\rm cri}$; see also \cite{Gouttenoire:2023naa}. These scaling behaviors correspond with our semianalytic analysis in Appendix~\ref{app:analytic}. 

Because of the exponential dependence of the PBH fraction on these input parameters, only a very narrow parameter region near the red contour with $f_{\rm PBH} = 1$ is practically viable to produce DM. On the other hand, from a phenomenological point of view, any reasonable PBH mass $M$ and fraction $f_{\rm PBH}$ can be generated in this scenario by changing the model parameters.

\section{Discussion}
\label{sec:discussion}

In this paper, we presented a thorough calculation of the PBH mass distribution function arising from vacuum energy decay in a FOPT. We examined the condition of Ref.~\cite{Liu:2021svg} for a Hubble volume to be entirely in the FV, and found that the surrounding wall cone was neglected. This omitted factor was found numerically to be of similar magnitude to the primary contribution of the Hubble volume itself and can therefore result in significant suppression of the PBH abundance.

Through our numerical simulations, we generated PBH populations that make up all of DM in the PBH mass window and that can originate the candidate HSC and OGLE microlensing events. Although we included the additional term and have a slightly different formalism, our results generally agreed with those of Ref.~\cite{Liu:2021svg}. Because of the sensitive dependence of the PBH abundance on the FOPT parameters, the extra suppression we found can be compensated by a modest change in parameters.

The PBH formation probability and the resulting abundance are exponentially sensitive to FOPT speed $\beta$ and strength $\alpha$. Among them, the dependence on $\beta$ was shown to be exceedingly strong, while the dependence on $\alpha$ was relatively mild, and further suppressed for $\alpha \gtrsim 10$. These are supported by semianalytic analysis, and they also align with intuition. For horizon-sized perturbations to survive late into the FOPT, small $\beta/H\lesssim \mathcal{O}(1)$ is needed as it determines the timescale of the phase transition and is inversely proportional to the bubble size~\cite{Kawana:2021tde,Lu:2022paj}. Naturally, small values of the vacuum energy suppress PBH formation and large values increase the formation rate, but make no further difference once a fully vacuum-dominated universe is reached.

Altogether, the PBH formation from horizon-sized vacuum decay is a viable and versatile scenario and can apply to a generic FOPT in beyond the Standard Model physics. Phenomenologically, setting the critical time/temperature determines the PBH mass, and minor changes in the FOPT parameters can produce the desired PBH mass function. 
We found that significant production of PBH occurs in phase transitions that are slow and moderately strong.

\section*{ACKNOWLEDGMENTS} 
The work of K. K. is supported by KIAS Individual Grants, Grant No. 090901.
The work of T. H. K. is supported by a KIAS Individual Grant No. PG095201 at Korea Institute for Advanced Study and National Research Foundation of Korea under Grant No. NRF-2019R1C1C1010050. 
The work of P. L. is supported by National Research Foundation of Korea under Grant No. NRF-2019R1C1C1010050. 
We thank Hyung Do Kim, Wan-Il Park, Dhong Yeon Cheong, Gansukh Tumurtushaa, Masahide Yamaguchi, Chang Sub Shin, and Alexander Kusenko for constructive comments which helped improve the paper.

\appendix

\section{Consistency check for completion of FOPT} \label{app:completion}

Here we show that our scenario is generically consistent with the completion of FOPT. We first discuss several options for completing a FOPT with (super)horizon-sized FV regions at $t_{\rm 1.45}$, and then present a numerical demonstration under a simplified treatment.

The major concern is that the horizon-sized FV regions soon become vacuum energy dominated and undergo a de Sitter expansion, starting a second inflation that resembles the old inflation scenario. Since the latter was abandoned due to the failure of the graceful exit, we must also check whether these FV regions do not harm the completion of FOPT. 

We first state that just having a de Sitter expanding FV region does not necessarily mean a never-ending inflation. The precise reason that the old inflation was unsuccessful is that the constant nucleation rate was restricted to be much smaller than $H^4$ in order to solve the horizon and the flatness problems. This turned out to be incompatible with the completion of the vacuum transition and thermalization \cite{Guth:1982pn}. Here in our scenario, no such restriction applies and we can have much higher nucleation rates during the phase transition. If the nucleation rate is high enough, the bubble nucleation and growth can overcome the volume expansion of FV so that the physical FV volume decreases to zero and the phase transition ends \cite{PhysRevD.46.2384,Kawana:2022fum}.

To clarify, assume a simplified situation of a de Sitter expanding background with constant $H$ and a constant nucleation rate $\Gamma$. The physical FV volume is $V^{\rm phys}_{\rm fv} \propto f_{\rm fv}(t) a^3(t)$, so its decrement requires
\begin{eqnarray}
    &&\frac{1}{V^{\rm phys}_{\rm fv}} \frac{d V^{\rm phys}_{\rm fv}}{dt} < 0  \nonumber \\
    && \quad = 3H(t) - \frac{4\pi}{3} \frac{d}{dt} \int^{t}_{t_{\rm cri}} dt' \, a^3(t') \, \Gamma (t') \, r^3(t, t') \qquad
\end{eqnarray}
which is, for constant $H$ and $\Gamma$,
\begin{equation}
    \Gamma > \frac{9H^4}{4\pi} \approx 0.7 H^4. \label{eq:GammadS}
\end{equation}
Thus, even a quite slow FOPT can complete while having horizon-sized FV regions if the nucleation rate in those regions is higher than about $0.7 H^4$.

Then the question comes down to how high the nucleation rates in the FV regions can be after passing $T_{\rm cri}$. This is determined by the temperature dependence of the effective potential as discussed in Sec.~\ref{sec:FOPT}. If the barrier disappears at some finite temperature, the nucleation rate in FV regions keeps increasing, and hence there is no problem in finishing the phase transition; after the barrier disappears, the phase transition becomes second order and ends through spinodal decomposition. On the other hand, if the barrier persists in the zero-temperature limit, the nucleation rate by thermal fluctuation decreases to zero. Then nucleation can only occur through quantum tunneling, and it is likely that the phase transition never ends, resulting in a second inflationary phase.

Typically, the latter case does not pose a serious problem because the remaining FV regions are fragmented into subhorizon sizes before the nucleation rate becomes too small. These regions are totally converted into TV by the bubble walls approaching from outside, instead of relying on bubble nucleations inside. However, as our scenario requires having horizon-sized FV regions at least until $t_{\rm 1.45}$, we need a stronger condition to have a consistent scenario.

In conclusion, there are generically three ways to have completion of FOPT within our scenario. First, if the effective potential gives the bubble nucleation rate that keeps increasing with decreasing temperature, the FOPT will finish without any problem. The exponential approximation we take \eqref{eq:expnucl} in our study corresponds to this case. Second, even if the nucleation rate peaks and decreases it is still possible to have a high enough nucleation rate after $t_{\rm 1.45}$ so that all the horizon-sized FV regions are dismantled into subhorizon pieces before the nucleation rate becomes too small. The third option is to have a sufficiently high quantum tunneling rate that can finish the FOPT even at zero temperature. 

The actual nucleation rates in these three cases will have deviations from the simple analytic exponential approximation in Eq.~(\ref{eq:expnucl}), although more realistic forms of the nucleation rate for the current scenario are left to future studies. 

In the rest of this appendix, we numerically demonstrate the decrease of physical FV volume after $t_{\rm 1.45}$ for the first option in a simplified setup. We regard the FV region at $t_{1.45}$ as a separate universe and assume it undergoes a de Sitter expansion with the Hubble rate coming from the vacuum energy density $\Delta V$. These are quite good approximations for horizon-sized FV regions, as they are vacuum dominated and the bubble walls from outside cannot entirely cover them. The exponential approximation (\ref{eq:expnucl}) is used for the nucleation rate.

\begin{figure}[t] \centering 
\includegraphics[width=0.48\textwidth]{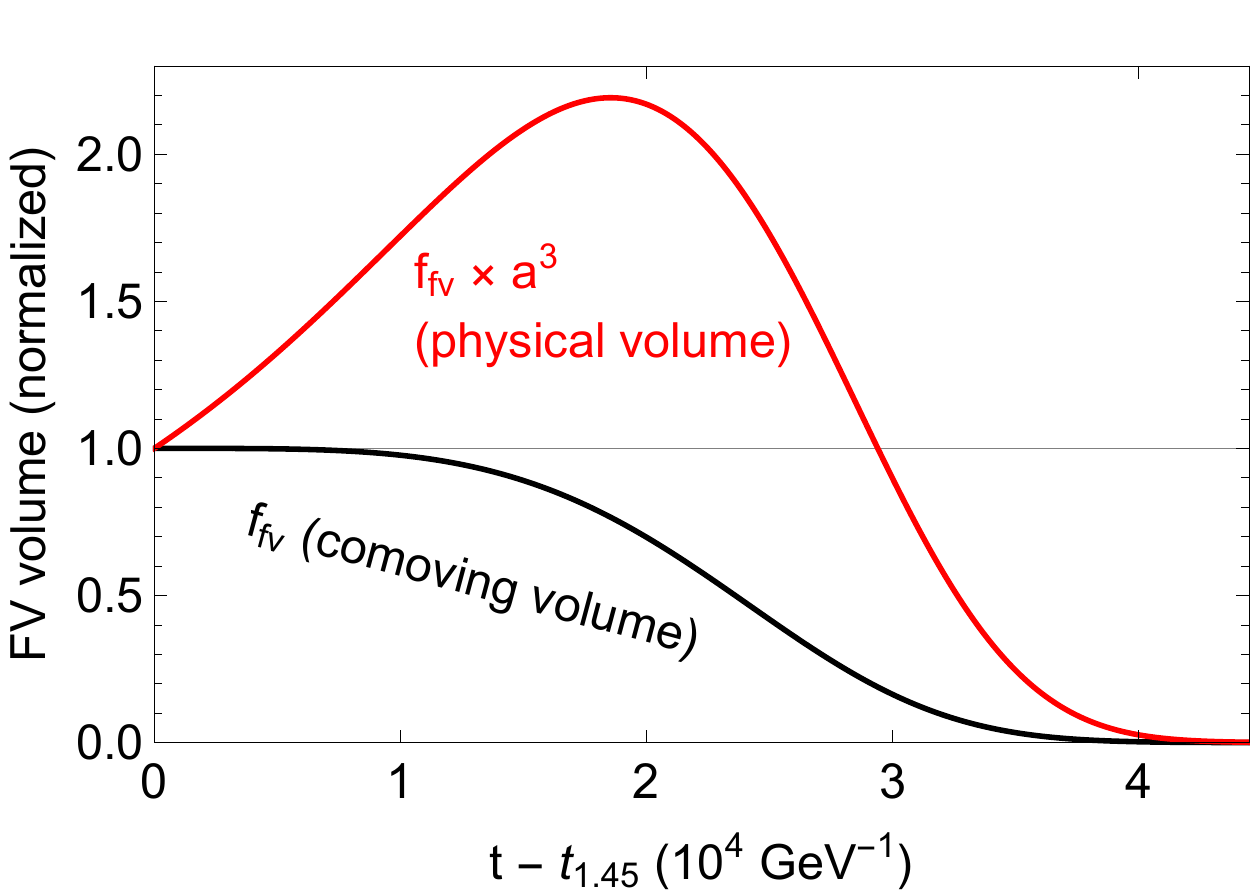}
\caption{
The normalized FV volume of a FV region at $t_{1.45}$, simplified to be a separated de Sitter universe. Black: the FV volume fraction equivalent to the normalized FV comoving volume. Red: the FV volume fraction times $a^3$ equivalent to the normalized FV physical volume.
} 
\label{fig:ffvs}
\end{figure}

In Fig.~\ref{fig:ffvs}, we plot $f_{\rm fv}(t)$ and $f_{\rm fv}(t) \, a^3(t)$ for the separated universe corresponding to a FV region at $t_{1.45}$ with the same parameters as in Fig.~\ref{fig:rhos}. We see that the physical volume of the FV initially increases due to the volume expansion, but it soon rapidly decreases due to the nucleation and growth of TV bubbles. The nucleation rate at $t_{1.45}$ is already about $\Gamma(t_{1.45}) \sim \mathcal{O}(10) \times H^4$ due to the exponential growth from $t_{\rm cri}$, and even keeps increasing further. Referring to Eq.~\eqref{eq:GammadS}, this nucleation rate is sufficient to convert all the remaining FV regions soon after the nucleation begins.

We also note that this large nucleation rate at late times is one of the two reasons why the survival probability [or $p(t)$ as its differential] is suppressed as depicted in Fig.~\ref{fig:p}. The other reason is of course the wall cone contribution in Fig.~\ref{fig:wallconeQ2Q3}. This tiny probability is compensated by the cosmological expansion during the radiation-dominated epoch, where the radiation density redshifts $\propto 1/a^4$ while the PBH only undergoes a number density dilution $\propto 1/a^3$. Because of this effect, the small fraction of horizons that have delayed transitions in the early Universe can contribute significantly to $\Omega_{\rm PBH}$ in the present day.

\section{Semianalytic analysis of PBH formation probability}
\label{app:analytic}
We derive the qualitative dependence of the PBH formation probability on the FOPT parameters $\alpha$ and $\beta$, and confirm the agreement to the intuitive expectations while also supporting our numerical results in Sec.~\ref{ssec:numerical}. As the actual PBH formation in this model is exponentially sensitive to the parameters, the quantitatively correct probabilities and resultant abundances need to be numerically computed as in the main text. 

Here we focus on the formation probability (Fig. \ref{fig:p}), as this is the main point of our calculation; analytic expressions of $\beta(M)$ and $f(M)$ follow directly from Eqs.~\eqref{eq:p},~\eqref{eq:betaM}, and~\eqref{eq:fPBH}. We approximate the integrated formation probability $\sim \int_{t_{1.45}} p(t) dt$ as simply the survival probability $P(t_{1.45})$, since the majority of the PBH production occurs at $t_{1.45}$ [see the rapidly dropping curves of $p(t)$ in Fig.~\ref{fig:p}]. Furthermore, it is sufficient to consider only the first exponent in Eq.~(\ref{eq:survival}) for our purposes since the behavior of the second exponent is similar (see Fig.~\ref{fig:wallconeQ2Q3}). 

As can be seen in Fig.~\ref{fig:rhos}, the average energy density $\Bar{\rho}$ stays approximately constant before $t_{1.45}$. This is generically true in our scenario, as the vacuum energy necessarily becomes dominant over the SM plasma. Then as the FV energy is converted into latent heat, the average density will decline to a value $\Bar{\rho}= (\Delta V + \rho_{\rm SM}(t_{1.45})) / 1.45 \approx 0.69 \Delta V$ by $t = t_{1.45}$. Thus, the average density and the Hubble rate [Eq.~\eqref{eq:Hubble}] are approximately constant for $t_{\rm cri} < t < t_{1.45}$ as $H_*^2 \simeq \Delta V / 3 M_P^2$, and hence, $a(t) \propto e^{H_* t}$.

Then in this constant Hubble rate approximation, the first exponent of Eq.~(\ref{eq:survival}) is (we calibrate $t$ by setting $t_{\rm cri} = 0$ hereafter)
\begin{eqnarray}
    \ln P_{\rm HV} (t_{1.45}) &\simeq& -\frac{4\pi}{3} \int^{t_{1.45}}_0 e^{-3H_* (t_{1.45} - t)} \frac{\Gamma_{\rm cri}}{H_*^3} \, e^{\beta t} \, dt \nonumber \\
    &\simeq& -\frac{4\pi}{3} \frac{\Gamma_{\rm cri}}{(3H_* + \beta) H_*^3} \, e^{\beta \, t_{1.45}} \label{eq:lnPHV}
\end{eqnarray}
where the subscript HV denotes the Hubble volume only (excluding the wall cone contribution) and $e^{\beta t_{\rm 1.45}}$, $ e^{3H t_{\rm 1.45}} \gg 1$. Since $H_{\ast}^2\propto \alpha$, only the behavior of $e^{\beta t_{1.45}}$ remains to be figured out.

To see how it scales, we use Eqs.~\eqref{eq:t145} and \eqref{eq:rhototavg}-\eqref{eq:rhoravg}. First, integrating Eq.~\eqref{eq:rhoravg} with Eqs.~\eqref{eq:rhovavg} and \eqref{eq:rhoR} gives
\begin{equation}
    \bar{\rho}_R(t) = \rho_{\rm SM} (0) e^{-4 H t} + \Delta V \int^t_{0} \left(-\frac{df_{\rm fv}}{dt'} \right) e^{-4 H_* (t - t')} dt'. \label{eq:rhoRintegrate}
\end{equation}
Then substituting Eqs.~\eqref{eq:rhototavg} and \eqref{eq:rhoRintegrate} into Eq.~\eqref{eq:t145} and eliminating $\Delta V$ by $\alpha \equiv \Delta V/\rho_{\rm SM}^{}(T_{\rm cri})$ gives
\begin{equation}
    f_{\rm fv} (t_{1.45}) + \frac{0.31}{\alpha} e^{-4 H_* \, t_{1.45}} + (1-f_{\rm fv} (t_{1.45})) X = 0.69~, \label{eq:rhoanal}
\end{equation}
where $X$ represents the average redshift of the transitioned vacuum energy defined by
\begin{equation}
    (1-f_{\rm fv} (t_{1.45})) X \equiv \int^{t_{1.45}}_{t_{\rm cri}=0} \left(- \frac{df_{\rm fv}}{dt} \right) e^{-4 H_* (t_{\rm 1.45} - t)} \, dt~.\label{eq:X}
\end{equation}
From the constant Hubble rate approximation in Eq.~\eqref{eq:ffvexp},
\begin{equation}
    f_{\rm fv}(t_{1.45}^{}) \simeq \exp\left[-\frac{8\pi v_w^3 \Gamma_{\rm cri} \, e^{\beta \, t_{1.45}^{}} }{\beta (H_*+\beta) (2H_*+\beta) (3H_*+\beta)} \right]. \label{eq:ffv145}
\end{equation}
From Eq.~\eqref{eq:rhoanal}, we extract the variation of $f_{\rm fv}(t_{1.45})$ with $\alpha$ and $\beta$. Then we use Eq.~\eqref{eq:ffv145} to solve for $e^{\beta t_{1.45}}$ as a function of $f_{\rm fv}(t_{1.45})$ and determine the behavior of $P_{\rm HV}$ in Eq.~\eqref{eq:lnPHV}.

We first examine the dependence on $\beta$. Faster phase transitions with larger $\beta$ have smaller $t_{1.45}$, which results in less redshifting and greater $X$.\footnote{While not mathematically clear, it is physically reasonable, and we also numerically checked for all three cases in Table \ref{tab:params}, with varying $\beta$ by $\mathcal{O}(1)$ times.}
The FV fraction at formation, $f_{\rm fv}(t_{1.45})$, decreases to satisfy Eq.~\eqref{eq:rhoanal} and lower the background energy density.\footnote{At linear order, the variation of the lhs in Eq.~(\ref{eq:rhoanal}) gives $(1-f_{\rm fv})\delta X+(1-X)\delta f_{\rm fv}^{}$. Since $1-f_{\rm fv}^{}\geq 0$, $\delta X>0$ means $\delta f_{\rm fv}^{}<0$ to maintain Eq.~(\ref{eq:rhoanal}); note that the second term containing $\alpha$ also increases.} Equation~\eqref{eq:ffv145} then demands $e^{\beta t_{1.45}}$ to greatly increase to get over the denominator. Solving for it gives the scaling of $e^{\beta t_{1.45}} \sim -\beta^4 \ln f_{\rm fv}$. Since $0 < f_{\rm fv} < 1$ itself decreases with increasing $\beta$, using this relation in Eq.~\eqref{eq:lnPHV} shows that $P_{\rm HV} (t_{1.45})$ decreases rapidly, faster than negative exponential of powers of $\beta$.

On the other hand, Eq.~\eqref{eq:rhoanal} shows that larger $\alpha$ results in larger $f_{\rm fv}(t_{1.45})$,\footnote{Similar to the previous case, this is not mathematically clear but physically reasonable since the critical density contrast will be reached in the earlier stage of FOPT. 
We checked this numerically too.} but the effect is suppressed by the exponential factor $e^{-4H_*^{}t_{1.45}^{}}$ and $\alpha$ itself. Furthermore, Eqs.~\eqref{eq:lnPHV} and \eqref{eq:ffv145} show no explicit $\alpha$ dependence as long as  $\Gamma_{\rm cri}/H_*^4$ and $\beta/H_*$ are fixed. Therefore, $P_{\rm HV} (t_{1.45})$ increases with $\alpha$, but the effect is suppressed if $\alpha$ keeps growing.

These behaviors are depicted in the bottom panel of Fig.~\ref{fig:p} and Fig.~\ref{fig:paramdep}. Increasing $\beta$ drastically lowers $p(t)$ and hence $f_{\rm PBH}$. Increasing $\alpha$ gives larger $p(t)$ and $f_{\rm PBH}$ but relatively mildly. Clearly, these agree with the rule of thumb expectations: The faster the phase transition, the less time for the redshift to generate density contrast, so PBH formation probability decreases with $\beta$; greater vacuum energy (the source of density contrast) over the homogeneous SM radiation always facilitates the PBH production, so the probability increases with $\alpha$, but asymptotes at large values of $\alpha$ corresponding to a vacuum-dominated universe.

\bibliography{references}

\end{document}